\documentclass[conference]{IEEEtran}

\usepackage{amsmath}
\usepackage{graphicx}
\usepackage{url}
\usepackage{amssymb}
\usepackage{adjustbox}
\usepackage{caption}
\usepackage{subcaption}
\usepackage{paralist}
\usepackage[ruled, vlined, linesnumbered]{algorithm2e}
\usepackage{algorithmic}
\usepackage{cite}

\newtheorem{defn}{Definition} 

\usepackage{makecell}

\usepackage{xcolor}
\newcommand{\planned}[1]{\textcolor{blue}{#1}}

\renewcommand{\planned}[1]{}

\newcommand{\fref}[1]{Fig.~\ref{#1}}

\title{A Reinforcement Approach for Detecting P2P Botnet Communities in Dynamic Communication Graphs}

\author{\IEEEauthorblockN{Harshvardhan P. Joshi and Rudra Dutta}
\IEEEauthorblockA{Department of Computer Science, 
North Carolina State University \\ 
Raleigh, NC 27695-8206, USA\\
Email: \{hpjoshi, rdutta\} @ncsu.edu}
}

\date{}

\begin{document}

\maketitle

\begin{abstract}
Peer-to-peer (P2P) botnets use decentralized command and control networks that make them resilient to disruptions. The P2P botnet overlay networks manifest structures in mutual-contact graphs, also called communication graphs, formed using network traffic information. It has been shown that these structures can be detected using community detection techniques from graph theory. These previous works, however, treat the communication graphs and the P2P botnet structures as static. In reality, communication graphs are dynamic as they represent the continuously changing network traffic flows. Similarly, the P2P botnets also evolve with time, as new bots join and existing bots leave either temporarily or permanently. In this paper we address the problem of detecting such evolving P2P botnet communities in dynamic communication graphs. We propose a reinforcement-based approach, suitable for large communication graphs, that improves precision and recall of P2P botnet community detection in dynamic communication graphs.

\end{abstract}

\section{Introduction}
Botnets are used for malicious purposes, such as spam and denial of service, with huge economic costs to the society. Decentralized command \& control structures of peer-to-peer (P2P) botnets make them more resilient to disruptions. However, these P2P overlay structures appear in communication graphs that are built from network flow meta-data, and can be detected using community detection techniques from graph theory. This is a promising approach for P2P botnet detection because it can work independent of device hardware and software, and is resilient to obfuscations employed by the botnets. 

\begin{figure}[tbp]
\centering
\includegraphics[width=0.95\columnwidth]{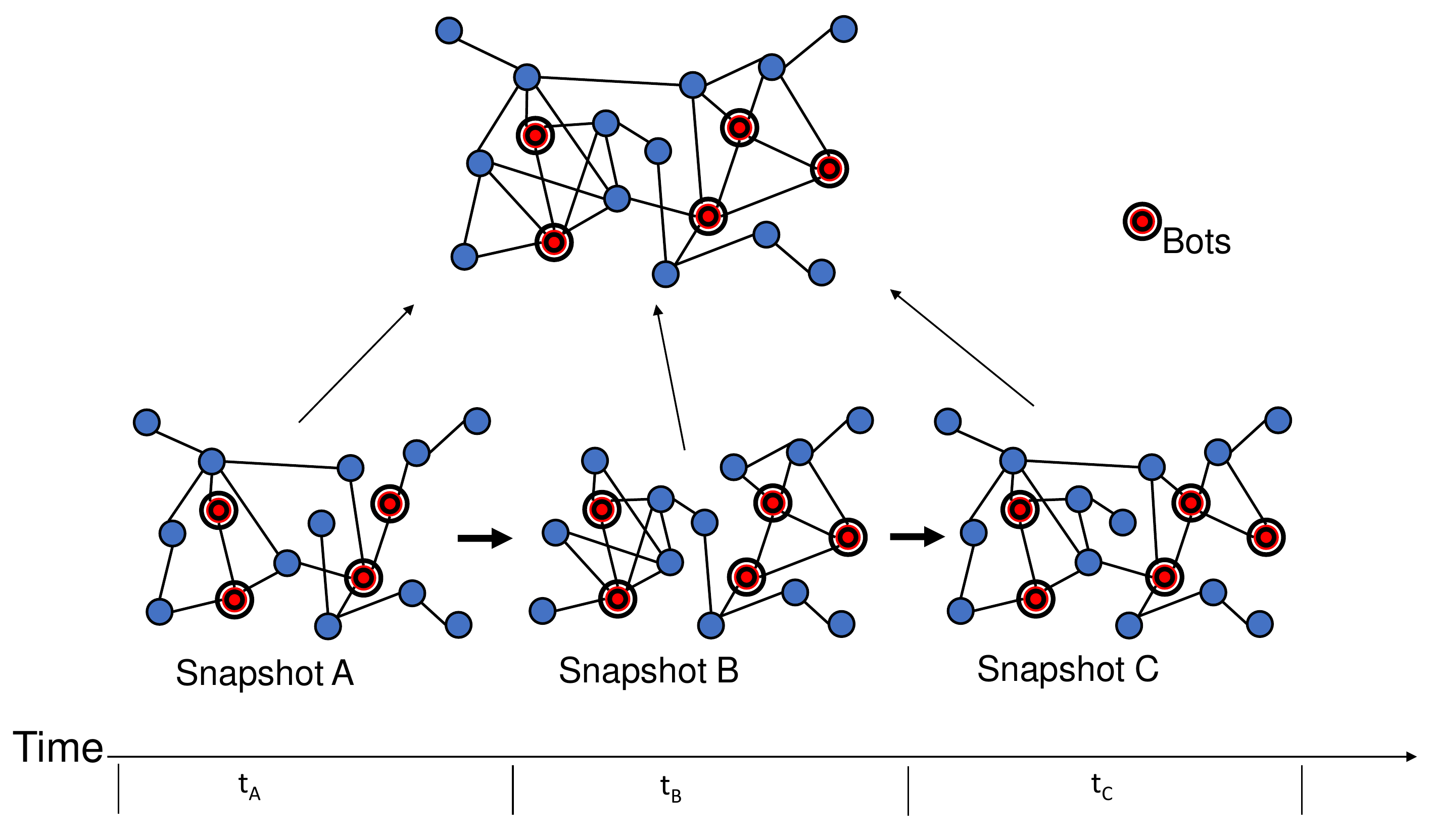}
\caption{Dynamic communication graphs and evolving P2P botnet communities. Each of the snapshot graphs associated with a time window contribute to the graph in the time frame of interest.}
\label{fig:evolvingBotnets}
\end{figure}

Several previous works have proposed various community detection based P2P botnet detection algorithms~\cite{zhuangPeerhunterDetectingPeertopeer2017, joshiImprovedP2PBotnet2019, nagarajaBotGrepFindingP2P2010, joshiGADFlyFastRobust2018, coskunFriendsEnemyIdentifying2010}. The communities detected by such algorithms correspond well with P2P botnets. However, these works assume a static communication graph, such as a snapshot graph built from network traffic flows observed within a window of time. Given that the network traffic flows continue outside that time window, any community structure analysis in another time window requires another graph snapshot. Along with the differences in connections (or edges) between nodes (or vertices) in these snapshot graphs across different time windows, new nodes will appear and existing nodes disappear as new communication endpoints become active or inactive in each time window, as shown in~\fref{fig:evolvingBotnets}.

How communities in a communication graph relate to each other across such sequence of snapshots is an open question. More directly relevant to the problem of P2P botnet detection, how P2P botnet communities relate to each other across communication graph snapshots can determine the effectiveness of a botnet community detection algorithm over a longer time period. These questions are addressed in this paper.

We make the following contributions in this paper:
\begin{itemize}
    \item formulate the problem of detecting evolving P2P botnets in dynamic communication graphs,
    \item propose schemes to leverage temporal information in the dynamic communication graphs based on reinforcing community memberships, and
    \item evaluate the proposed schemes with dynamic botnet and communication graphs built from real-world traffic data.
\end{itemize}

In addition, while several works have proposed to use temporal characteristics of traffic like flow-duration, periodicity, or inter-packet arrival times to detect P2P botnet traffic, such characteristics can potentially be obfuscated by botnets to avoid detection. Our proposed approach instead uses a coarser temporal trend, that the P2P botnet community is likely to be more stable over time since it needs to maintain connectivity through such ad-hoc overlay network.

\section{Background and Related Work}
There have been several works on botnet detection using botnet communication patterns, including the temporal information such as inter-packet interval and the periodicity and duration of flows. However, to the best of our knowledge, this is the first work that addresses the P2P botnet community detection in dynamic communication graphs. 

There have been several works in the graph theory and complex networks fields on community detection in dynamic graphs. These works usually focus on characterizing or tracking communities as they evolve. A comprehensive overview and survey of community detection in graphs, including in dynamic graphs, has been presented by S. Fortunato~\cite{fortunatoCommunityDetectionGraphs2010}. An online graph clustering algorithm is presented by Zanghi et al.~\cite{zanghiFastOnlineGraph2008}. This algorithm however has time complexity of $O(N^2)$, and thus is not suitable for very large graphs with millions of vertices. Multi-layered graphs can be used to represent dynamic or temporally changing graphs. The stochastic block models have been used to infer community structure in multi-layered graphs including dynamic graphs by T. P. Peixoto~\cite{peixotoInferringMesoscaleStructure2015}. 
Community detection in multi-layered dynamic graphs using a modularity optimization approach is presented by Mucha et al.~\cite{muchaCommunityStructureTimeDependent2010}. Suitability of these methods for P2P botnet detection in very large communication graphs has not been studied.

\section{Problem Formulation}
\label{sec:DynProbForm}
In this section, we formalize the problem of detecting evolving P2P botnets in dynamic communication graphs. First, we define a model of discrete time intervals for capturing temporal changes to communication graphs. Then, P2P botnet detection is formally defined for a single communication graph, and for a sequence of dynamic communication graphs.

\subsection{Discrete Time Model of Dynamic Communication Graphs}

\begin{figure}[tbp]
\centering
\includegraphics[width=0.95\columnwidth]{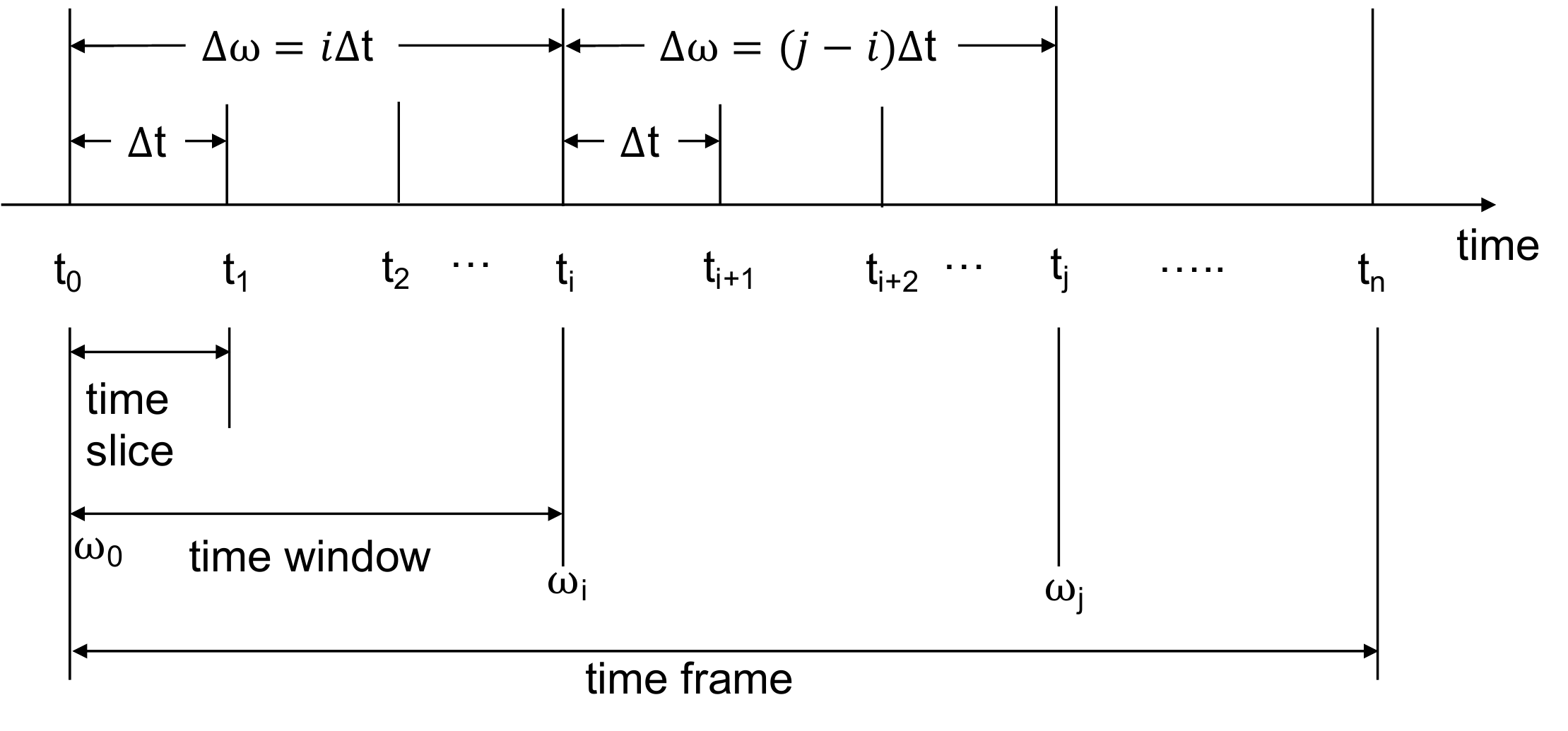}
\caption{Discrete time model for analyzing dynamic communication graphs.}
\label{fig:timeModel}
\end{figure}

In order to simplify the continuous changes in communication traffic flows on a continuous time scale, we propose a discrete time model of dynamic changes to communication graphs, shown in~\fref{fig:timeModel} and define some relevant terms related to dynamic communication graphs.

\begin{defn}
A \textbf{time slice} $t_i$ is an indivisible period of time $\Delta t$ in which communication traffic flows are observed. All communication flows within a single time slice is assigned the same temporal value $t_i$. 
\end{defn}

A time slice is uniquely associated with a temporal value on communication traffic flows, and the resulting edges in a communication graph. For consistency, we use the time at the start of a time slice as the temporal value associated with the time slice. The discrete duration of time $\Delta t$ is the difference between the temporal values of two contiguous time slices, i.e., $t_{i+1} - t_i = \Delta t$. Unless stated otherwise, each time slice is of the same time duration $\Delta t$.

\begin{defn}
A \textbf{time window} $\omega_m$ is a period of time consisting of one or more contiguous time slices ${t_m, t_{m+1}, \dots, t_{n-2}, t_{n-1}}$ in which communication traffic flows are observed. 
\end{defn}

The duration $\Delta \omega$ of the time window $\omega_m$ is the difference between the temporal values of its first time slice and the first time slice of the next time window. That is, $\omega_n - \omega_m = t_n - t_m = \Delta \omega$. Given time slices of equal duration $\Delta t$, the time window length, as a discrete measure of its duration, is given by the number of time slices within the time window, or $|\Delta \omega| = n - m$.


\begin{defn}
A \textbf{time frame} $T_i$ is a period of time, consisting of one or more contiguous time windows $\omega_i, \omega_{i+1}, \dots, \omega_{j-2}, \omega_{j-1}$, that is of interest for P2P botnet detection. 
\end{defn}

We limit our inquiry into the composition of the P2P botnet to a predefined time frame, and for the purposes of this discussion, the structures of P2P botnets in two distinct time frames are considered to be independent. 

\subsection{P2P Botnet Detection in Dynamic Communication Graphs}
\label{sec:DynProbFormulationP2P}

Next, we define P2P botnet community detection in the context of evolving botnet community and dynamic communication graphs. 

In a time window $\omega_i$, and the corresponding communication graph $G_i = (V_i, E_i)$, the P2P botnet C\&C communication manifests as a subgraph $G_{P_i} = (V_{P_i}, E_{P_i})$, such that, $V_{P_i} \subseteq V_i$ and $E_{P_i} \subseteq E_i$. The edges $E_{P_i}$ represent the P2P botnet C\&C traffic observable during the time window $\omega_i$, and the vertices $V_{P_i}$ represent the P2P bots participating in the P2P C\&C traffic during that time window. Thus, the subgraph $G_{P_i}$ is an edge induced subgraph of the communication graph $G_i$ from the edges $E_{P_i}$ corresponding to the P2P C\&C traffic during the time window $\omega_i$. 

P2P botnet community detection is a partitioning $C_i$, also called a set of communities, of a given communication graph $G_i$ associated with the time window $\omega_i$, and containing the P2P botnet subgraph $G_{P_i}$. In the context of P2P botnet detection, the partition $C_i = \{c_{1_i}, c_{2_i},\dots, c_{k_i}\}$ is an ordered set, with communities $c_{1_i},\dots, c_{k_i}$ ordered by the number of P2P botnet vertices they contain, i.e., $c_{r_i} \leq c_{s_i}$ if and only if $|c_{r_i} \cap V_{P_i}| \leq |c_{s_i} \cap V_{P_i}|$. The P2P botnet community is the community $c_{P_i}$ that is the maximum set of the partition $C_i$, or, $|c_{P_i} \cap V_{P_i}| \geq |c_{r_i} \cap V_{P_i}|, \forall c_{r_i} \in C_i$. 

In a time frame $T_i$, consisting of $n$ time windows, there are $n$ communication graphs $G_i, G_{i+1}, \dots, G_{n-1}$, with the corresponding P2P botnet subgraphs $G_{P_i}, G_{P_{i+1}}, \dots, G_{P_{n-1}}$. The set of P2P bots for this time frame is the union of all P2P bots from each time window of the time frame, i.e., $V_{P_{T_i}} = \bigcup_{j = i}^{n-1} V_{P_j}$. Note that the P2P botnet nodes $V_{P_{T_i}}$ are a subset of all the network nodes visible within the time frame $V_{T_i}$.

The problem of detecting P2P botnet communities in dynamic communication graphs can be defined as partitioning the set of visible network nodes $V_{T_i}$ such that the P2P botnet community $c_{P_{T_i}}$ is the maximum subset of the P2P botnet nodes $V_{P_{T_i}}$.

This paper addresses the problem of how to partition the communication graphs in each time window into partitions $C_i, C_{i+1}, \dots, C_{n-1}$, such that the corresponding P2P botnet communities $c_{P_i}, c_{P_{i+1}}, \dots, c_{P_{n-1}}$ are maximum size subsets of P2P botnet nodes $V_{P_i}, V_{P_{i+1}}, \dots, V_{P_{n-1}}$.

\section{Using Temporal Information in Dynamic Graphs}
\subsection{The Naive Approach}
A simple, and perhaps naive, approach to the analysis of communication graphs from dynamic network communication traffic is to divide the time frame of interest into disjoint and independent time windows, ignoring the temporal information available within the time window. We call the graphs following this approach as snapshot graphs, defined as
\begin{defn}
A \textbf{snapshot graph} $G_i = (V_i, E_i)$ is a simple and undirected communication graph that represents the communication flows within a single time window $\omega_i$. 
\end{defn}

A simple graph is a graph with no self loops, i.e., edges with a single vertex as both its endpoints, or multiple edges, i.e., more than one edge between any pair of vertices. A snapshot graph does not distinguish between edges contributed by individual time slices $t_i \dots t_j$ within the time window.
The communication graphs used in previous works such as~\cite{zhuangPeerhunterDetectingPeertopeer2017, joshiImprovedP2PBotnet2019, nagarajaBotGrepFindingP2P2010, joshiGADFlyFastRobust2018, coskunFriendsEnemyIdentifying2010} are snapshot graphs. In addition, they do not have any edge attributes with temporal information. The previous works have shown that P2P botnet community detection works well in such snapshot graphs. However, it is not clear that this snapshot-based approach is the best approach at P2P botnet community detection, within the time window given temporal information available for each time slice, or for the entire time frame. 

\subsection{Reinforcement Approach}
Irrespective of how well the naive approach detects the evolving P2P botnet communities, it certainly ignores the temporal information available within each time window. Since the temporal information relates to the traffic exchanged between network nodes, it can be added as an attribute to the edges in the communication graphs. There are several ways to embed the temporal traffic information as an edge attribute, for example, as the duration of the underlying network flow that the edge represents, or the sum of duration of all active flows between the node pair within the time window, or as a sequence of time intervals during with the flow is active. 

\subsubsection{Using Temporal Edge Weights}
A simple improvement to the naive approach is to assign a weight $w_{e_i}$ to the edges $e_i$ of the snapshot graph such that an edge has more weight if it appears in more time slices within the given window. 
\begin{defn}
A \textbf{composite graph} is an undirected communication multi-graph that represents the communication flows within a single time window, that is a union of snapshot graphs associated with each time slice in the window.
\end{defn}
This multi-graph, i.e. a graph with multiple edges between a vertex pair, can be simplified by combining multiple edges between a vertex pair into a single edge, with the sum of weights of original edges as the weight of the combined edge. Formally, the weight $w_{ij}$ of an edge $e_{ij} = (v_i, v_j)$ connecting vertices $v_i$ and $v_j$ in the composite graph $G_m = (V_m, E_m)$ for time window $\omega_m$, is the sum of weights of all edges between $v_i$ and $v_j$ in the constituent time slices $t_m, t_{m+1}, \dots, t_{n-1}$. 

The previously proposed community detection algorithms can be used on such composite graphs with edge weights, as the algorithms either incorporate edge weights for community detection, or they can be modified to use edge weights. 
The composite graph gives more structural weight to long-running traffic flows, or contact between node pairs that communicate with each other frequently over longer periods of time.

\subsubsection{Using Community Membership}
The consensus clustering approach proposed by Lancichinetti and Fortunato~\cite{lancichinettiConsensusClusteringComplex2012} uses the community membership information collected repeatedly, over the same time slice or across time slices, to build a consensus matrix as a representation of community co-membership, which in turn becomes an adjacency matrix of a consensus graph. The communities identified in this consensus graph can be considered to be the communities of the original dynamic graphs. However, due to the dense structure of the consensus graphs they can no longer be represented efficiently as adjacency list but rather as an adjacency matrix. In addition, the community detection on these denser graphs (large $|E|$) is significantly slower as even the faster community detection methods have a complexity in the order of $O(|E|log|V|)$.
Hence, this approach is not suitable for very large graphs with millions of vertices. 

Instead, we propose an approach inspired by the consensus clustering that is more suitable for very large graphs. The communication membership information from snapshot graphs of individual times slices can be used to reinforce existing edges in the communication graphs, by building a reinforced graph as below:
\begin{itemize}
    \item Partition the time slice graph $G_{t_i}$ into communities $C_{t_i}$, for each time slice $t_i$ in the time window $\omega_j$
    \item Increase the weight of community internal edges (edges with both vertices in the same community) in each time slice graph by a reinforcement factor $\gamma$.
    \item Build a composite graph $G_j$ for the time window $\omega_j$ using these reinforced time slice graphs.
\end{itemize}

\begin{defn}
A \textbf{reinforced graph} is an undirected communication multi-graph that represents the communication flows within a single time window, whose edges are reinforced by the community membership information from each of the time slices in the window.
\end{defn}
Again, just like the composite graph, multiple edges between a pair of vertices in the reinforced graph are simplified such that the combined edge has a weight equal to the sum of weights of the multiple edges. The community detection is repeated for this reinforced graph and the resultant communities are considered to be the communities for that time window.

The communities detected using the reinforced graph are likely to be communities that manifest their community structure across various time slices in the time window. Since we hypothesize that the P2P botnet communities are likely to be more stable than other communities in the communication graphs, the reinforcement approach outlined here is likely to improve the accuracy of detecting P2P botnets. 
Our proposed approach keeps the reinforced graph sparse, and thus uses more memory efficient adjacency list representation of the graph and also results in faster community detection.

\section{Evaluation}

\subsection{Experiment Methodology}

Using an initial seed list of peers of a botnet and the reverse engineered P2P protocol for that botnet, the P2P botnet can be crawled by repeatedly requesting a list of neighbors or peers from each peer until no new peers are found. This information can be used to reconstruct the topology of the P2P overlay network used by the botnet. We use a dataset of P2P graphs reconstructed from crawling the Sality P2P botnet~\cite{haasResilienceP2PbasedBotnet2016}. The data contains snapshots of reconstructed Sality P2P network graphs based on hourly crawling data over a period of 24 hours. These hourly P2P botnet graphs are considered to be the snapshot graphs for hourly time slices.

For background communication traffic data, we use passive monitored network traffic traces collected by the MAWI project in Japan~\cite{choTrafficDataRepository2000}. The network traffic traces collected through passive monitoring from this backbone link are available from 2006 through 2020 (present). The traces are collected daily for about 15 minutes at 1400 (i.e. 2 pm). Longer traces of 24 hours, 48 hours, 72 hours, and 96 hours are also available for certain periods, including as part of the A Day in the Life of the Internet project\cite{analysisDayLifeInternet}. We select a set of traces collected over a period of 24 hours in April 2019, from samplepoint-G which is the link connecting the MAWI network to an Internet Exchange~\cite{MAWIWorkingGroup}. Due to the volume of traffic in these traces, we use a 15 minute time slice of the traffic to create a snapshot communication graph that is combined with the hourly P2P botnet graphs.

\subsection{Planting Dynamic P2P Botnets in Dynamic Communication Graphs}
There are several ways of planting the dynamic P2P botnet graphs in dynamic background communication graphs. Given snapshots of P2P botnet graphs $G_{P_i} = (V_{P_i}, E_{P_i})$, with $i = 1, 2,\dots, N_P$ a set of all bots observed across the snapshots is $V_{P_{all}} = \bigcup_{i=1}^{N_P} V_{P_i}$, if each vertex is identified uniquely with a bot such as using its IP address. Similarly, the set of all nodes observed across the background communication graph snapshots $G_{C_i} = (V_{C_i}, E_{C_i})$, where $i = 1, 2,\dots, N_C$, is $V_{C_{all}} = \bigcup_{i=1}^{N_C} V_{C_i}$, where the vertex is again uniquely identified with an IP address. 

\subsubsection{How to map botnet nodes to communication graph nodes?}
A simple approach is to randomly map each P2P botnet node to a background communication graph node uniquely identified by its IP address, with mapping $f: V_{P_{all}} \to V_{C_{all}}$, since the number of bots are strictly less than the total number of observed nodes in the communication graphs $|V_{P_{all}}| < |V_{C_{all}}|$. There are several reasons however that this approach is not very realistic. 
\begin{itemize}
    \item The communication graph is generated from monitored traffic from a particular network, and hence some of the vertices in the communication graphs are internal nodes of the network while others are external nodes. Thus, traffic from these internal nodes is more observable than the external nodes.
    \item The communication graphs have power-law like degree distribution~\cite{joshiImprovedP2PBotnet2019}. Randomly selecting from all vertices is likelier to choose vertices with low degree, making it easier to detect higher degree botnet nodes. 
\end{itemize}

Since the communication graph is usually generated from monitored traffic from a particular network, it is more realistic to randomly map each sampled P2P botnet node to the monitored nodes in the background traffic. The traffic traces used in our experiments are anonymized and do not contain information on which nodes are internal to the network, so we use a proxy measure to identify such nodes. We note that the internal nodes of the network are likelier to appear repeatedly over longer periods of time in the traffic traces compared to most of the external nodes, though with some exceptions, such as commonly used services like Google or Facebook. Thus, we create a list of nodes that are active in each of the time slices within our time frame of interest, called the always active nodes, and map the botnet graph nodes to these active nodes.

A random sample of P2P botnet nodes are selected as monitored nodes whose traffic is visible as edges in the communication graphs. Since these nodes communicate with other P2P botnet nodes that are not monitored, the visible traffic includes those other P2P nodes as well. In one scenario, that is the most realistic, these monitored P2P botnet nodes are mapped to the always active nodes in the traffic traces, as a proxy to the monitored internal nodes of the network. The scenario represents the case where all botnet traffic of the internal nodes is observable but the botnet traffic of external nodes is only observable to the extent that they communicate with the internal botnet nodes. In another scenario that we believe represents a worst case, all P2P botnet nodes are mapped to the always active nodes. This represents the case where some of the monitored nodes' botnet traffic is not visible in the traffic traces.

\subsection{Effectiveness of Reinforcement Approach}
The effectiveness of the three schemes, naive and the proposed composite and reinforced community detection, are investigated in this section.

\subsubsection{Effectiveness Within a Time Window}
\begin{figure}[htbp]
\centering

	\begin{subfigure}[b]{0.49\columnwidth}
	\centering
	\includegraphics[width=\columnwidth]{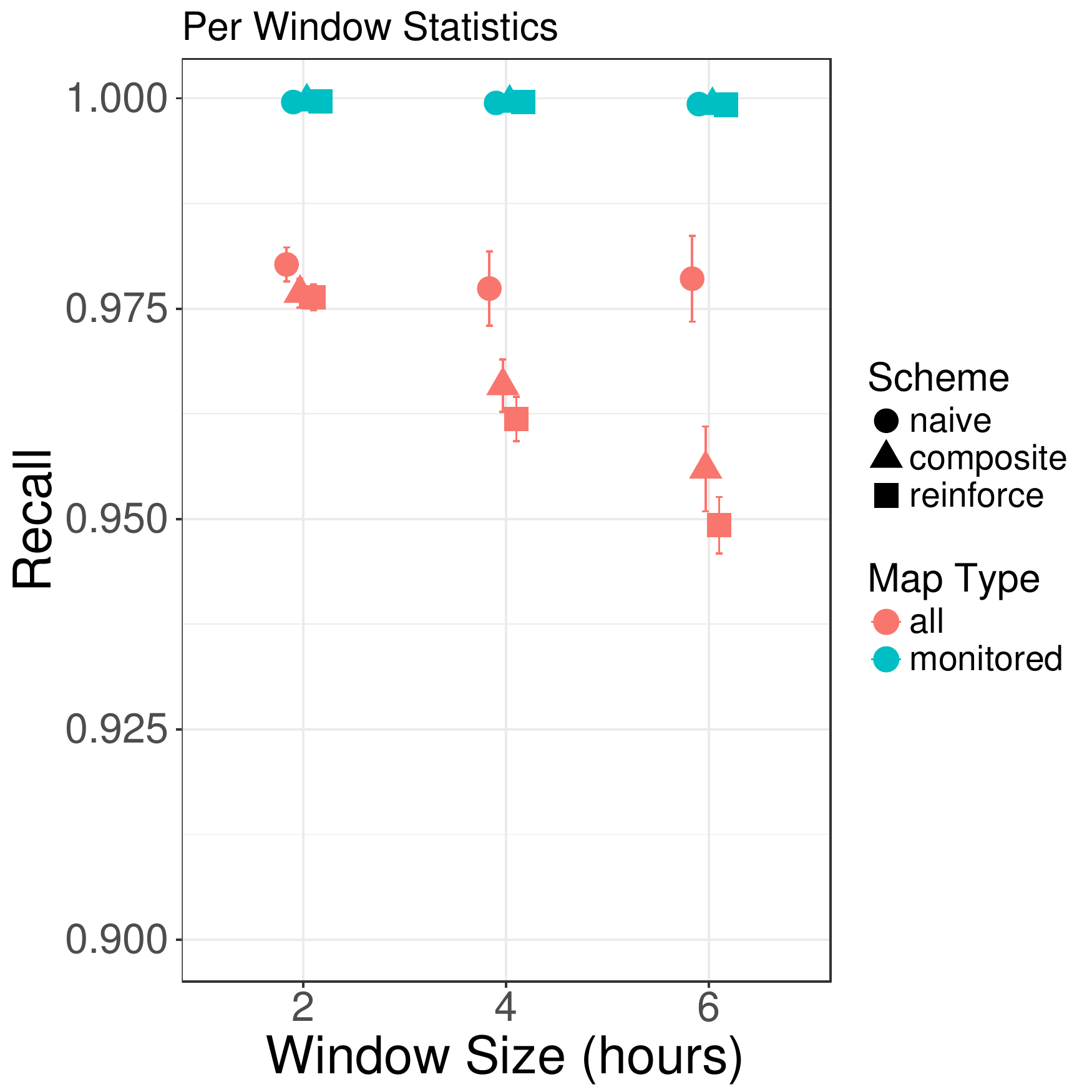}
	\caption{Recall}
	\label{fig:perWindowRecall}
	\end{subfigure}%
	\hfill%
	\begin{subfigure}[b]{0.49\columnwidth}
	\centering
	\includegraphics[width=\columnwidth]{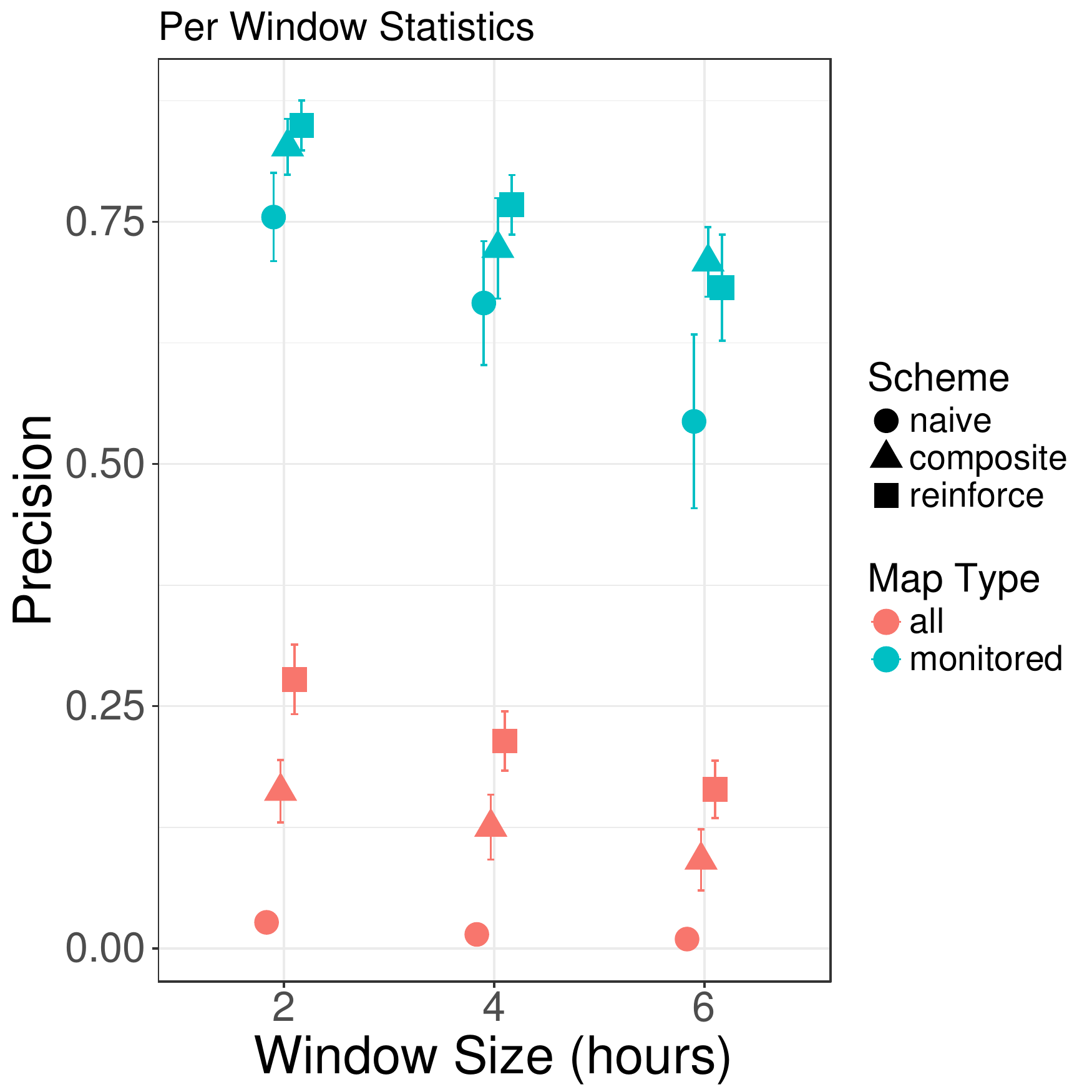}
	\caption{Precision}
	\label{fig:perWindowPrecision}
	\end{subfigure}
	\caption{Recall and Precision for different window sizes with the Louvain algorithm and the reinforcement schemes. The error bars represent 95\% confidence interval.}
\label{fig:perWindowRecallPrecision}
\end{figure}

We first ask how effective are the various schemes of dynamic P2P botnet detection within a given time window. In~\fref{fig:perWindowRecallPrecision} the recall and precision of P2P botnet detection with the Louvain community detection algorithm~\cite{blondelFastUnfoldingCommunities2008} is shown for different window sizes, and the three schemes proposed here for P2P botnet community detection in dynamic graphs. Each statistic is reported for two scenarios: when \emph{all} the actual P2P botnet nodes are mapped to any of the always active nodes in the traffic traces, and when only the randomly selected \emph{monitored} actual P2P nodes are mapped to any of the always active nodes in the traffic traces. 
For the worst case scenario of all P2P botnet nodes mapped to the always active nodes, the naive scheme has consistently close to 98\% recall, which is significantly higher than the close to 95\% recall with the reinforced community detection scheme for larger window sizes. However, the precision for the naive scheme is extremely low, with tens of thousands of false positives so that some of the P2P botnet nodes missed by the reinforcement approach are included in the detected P2P community by the naive approach. It is preferable to trade a few more false negatives from the reinforcement approach for the reduction in tens of thousands of false positives.
In the more realistic scenario of only monitored P2P botnet nodes being mapped to the always active nodes, the recall with all three schemes is nearly complete. However, the precision of reinforced community detection is significantly improved compared to the naive approach. We also note that as the window size increases, both the recall and precision reduces since some of the P2P botnet nodes do not stay active over the entire window while connections are accreted to other non-P2P nodes.

\begin{figure}[htbp]
\centering

	\begin{subfigure}[b]{0.49\columnwidth}
	\centering
	\includegraphics[width=\columnwidth]{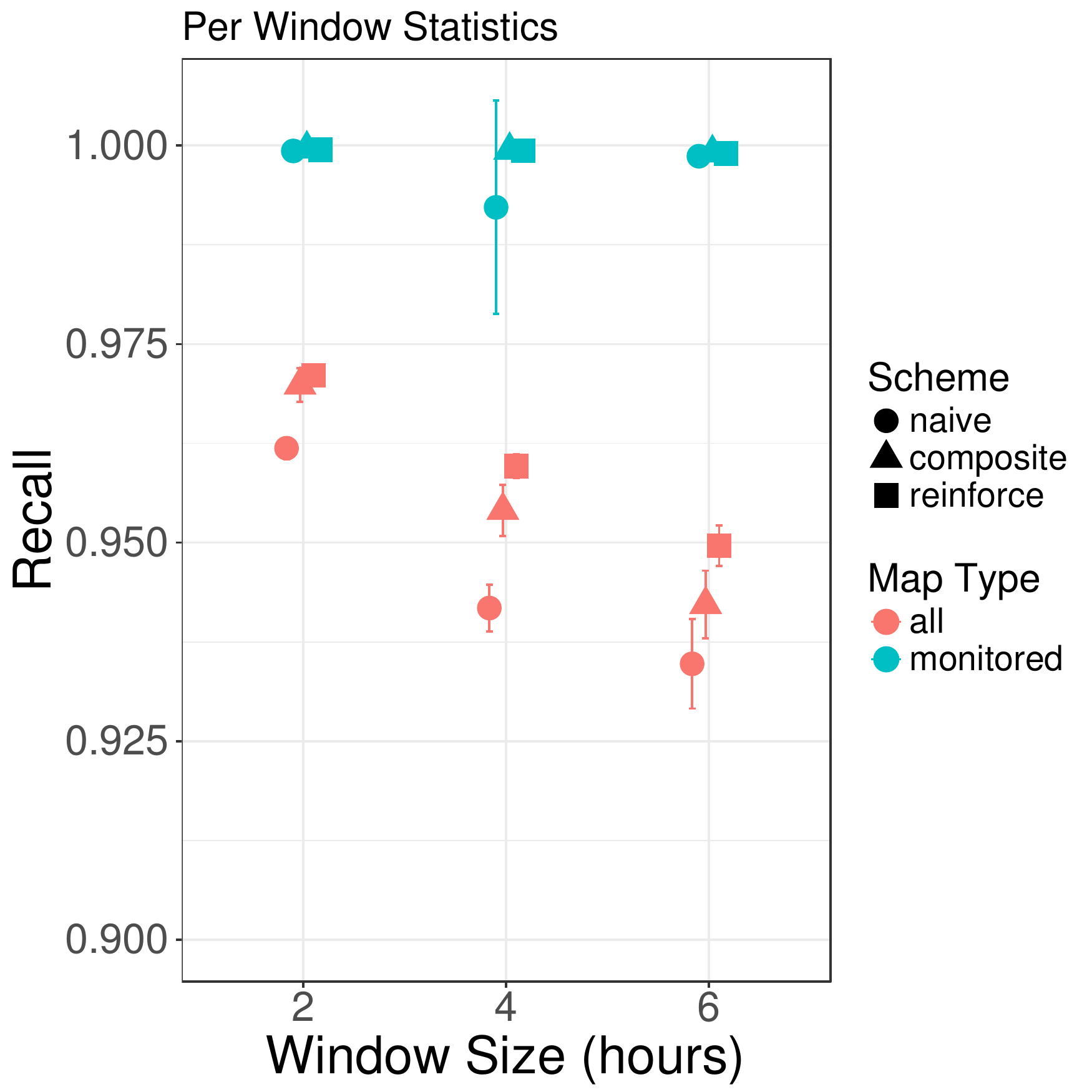}
	\caption{Recall}
	\label{fig:perWindowRecallLPA}
	\end{subfigure}%
	\hfill%
	\begin{subfigure}[b]{0.49\columnwidth}
	\centering
	\includegraphics[width=\columnwidth]{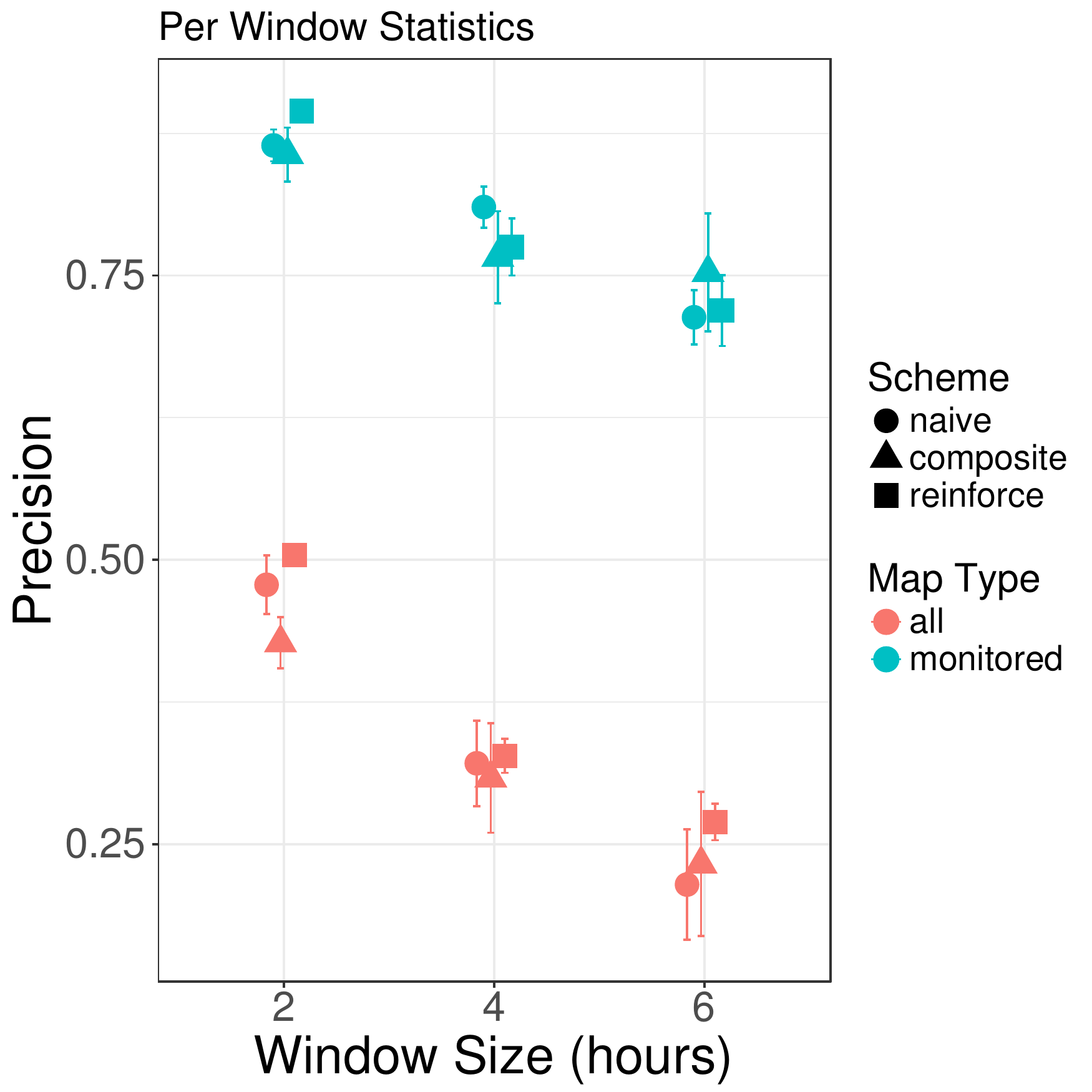}
	\caption{Precision}
	\label{fig:perWindowPrecisionLPA}
	\end{subfigure}
	\caption{Recall and Precision for different window sizes with the Label Propagation algorithm and the reinforcement schemes. The error bars represent 95\% confidence interval.}
\label{fig:perWindowRecallPrecisionLPA}
\end{figure}

Similarly, recall and precision for the Label Propagation algorithm (LPA)~\cite{raghavanLinearTimeAlgorithm2007} is presented in~\fref{fig:perWindowRecallPrecisionLPA}. The LPA has precision several times that of the Louvain algorithm for the worst case where all P2P botnet nodes are mapped to the always active nodes in the background traffic. As discussed in paper~\cite{joshiImprovedP2PBotnet2019}, LPA tends to detect smaller communities which results in much higher precision compared to the Louvain algorithm for P2P botnet community detection. On the other hand, LPA tends to have lower recall due to leaving out some less well connected P2P nodes, though in these experiments we see only slightly lower recall than the Louvain algorithm for the naive approach. However, the reinforcement-based schemes consistently improve the naive LPA recall, while also improving or maintaining the precision. 

Thus, across various window sizes, algorithms, and models for planting P2P botnet graphs in background communication graphs, we see that the reinforced community detection scheme improves precision of community detection within a time window. In addition, for algorithms like LPA that are less deterministic, the repeated community detection of reinforcement approach helps improve the recall as well.

\subsection{Effectiveness Across Time Windows}
Perhaps more relevant to P2P botnet detection in dynamic communication graph is the effectiveness of various approaches across time windows, within a time frame of interest. Since any network node participating in the P2P botnet at any time during the time frame is considered a part of the P2P botnet, we take a similar approach with the detected P2P botnet communities. As described in section~\ref{sec:DynProbFormulationP2P}, the predicted P2P botnet community for the time frame is the union of P2P botnet communities for the constituent time windows. It should be noted that this approach is feasible in our experiments since there is only one P2P botnet community of interest.

\begin{figure}[htbp]
\centering

	\begin{subfigure}[b]{0.49\columnwidth}
	\centering
	\includegraphics[width=\columnwidth]{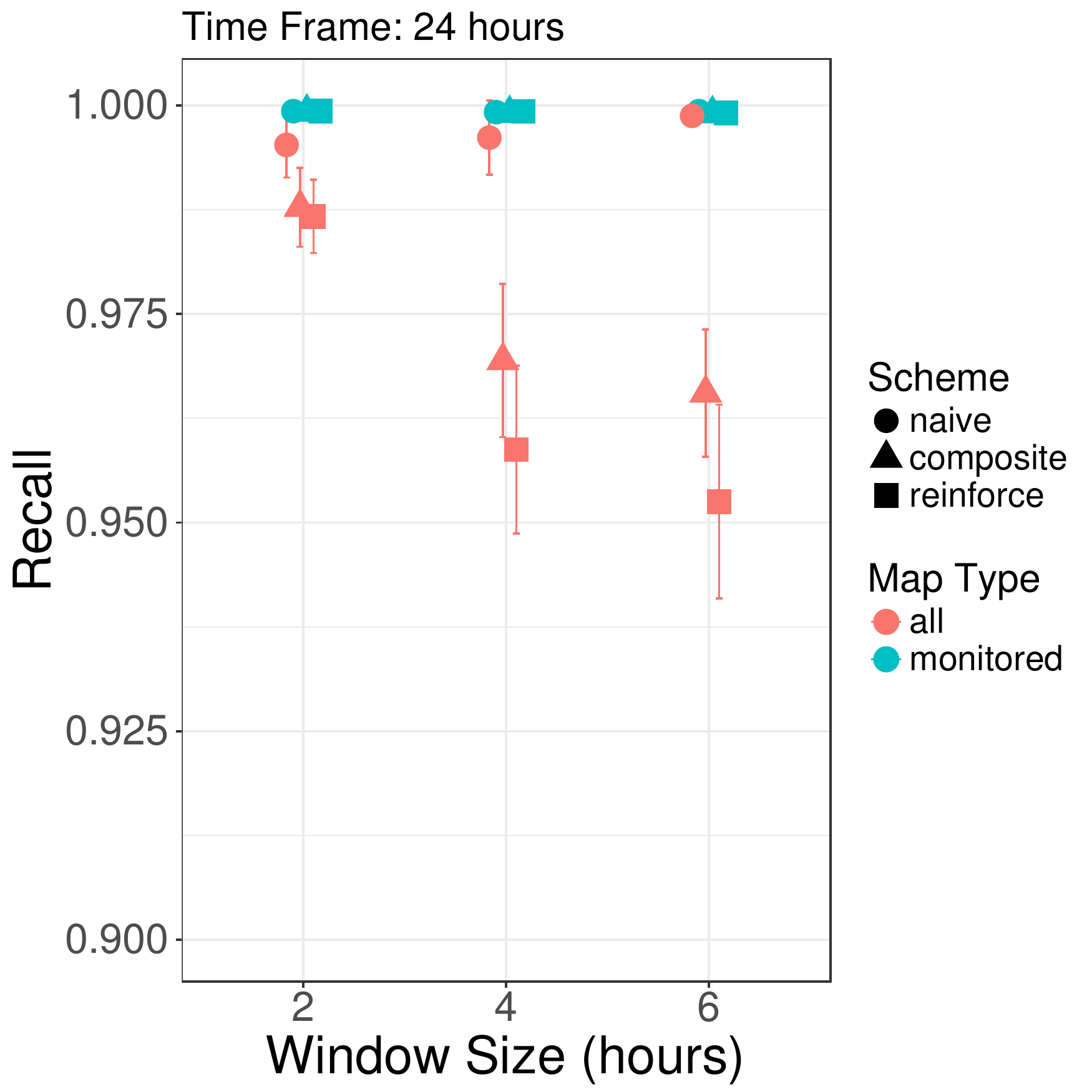}
	\caption{Recall}
	\label{fig:overallRecallMLA}
	\end{subfigure}%
	\hfill%
	\begin{subfigure}[b]{0.49\columnwidth}
	\centering
	\includegraphics[width=\columnwidth]{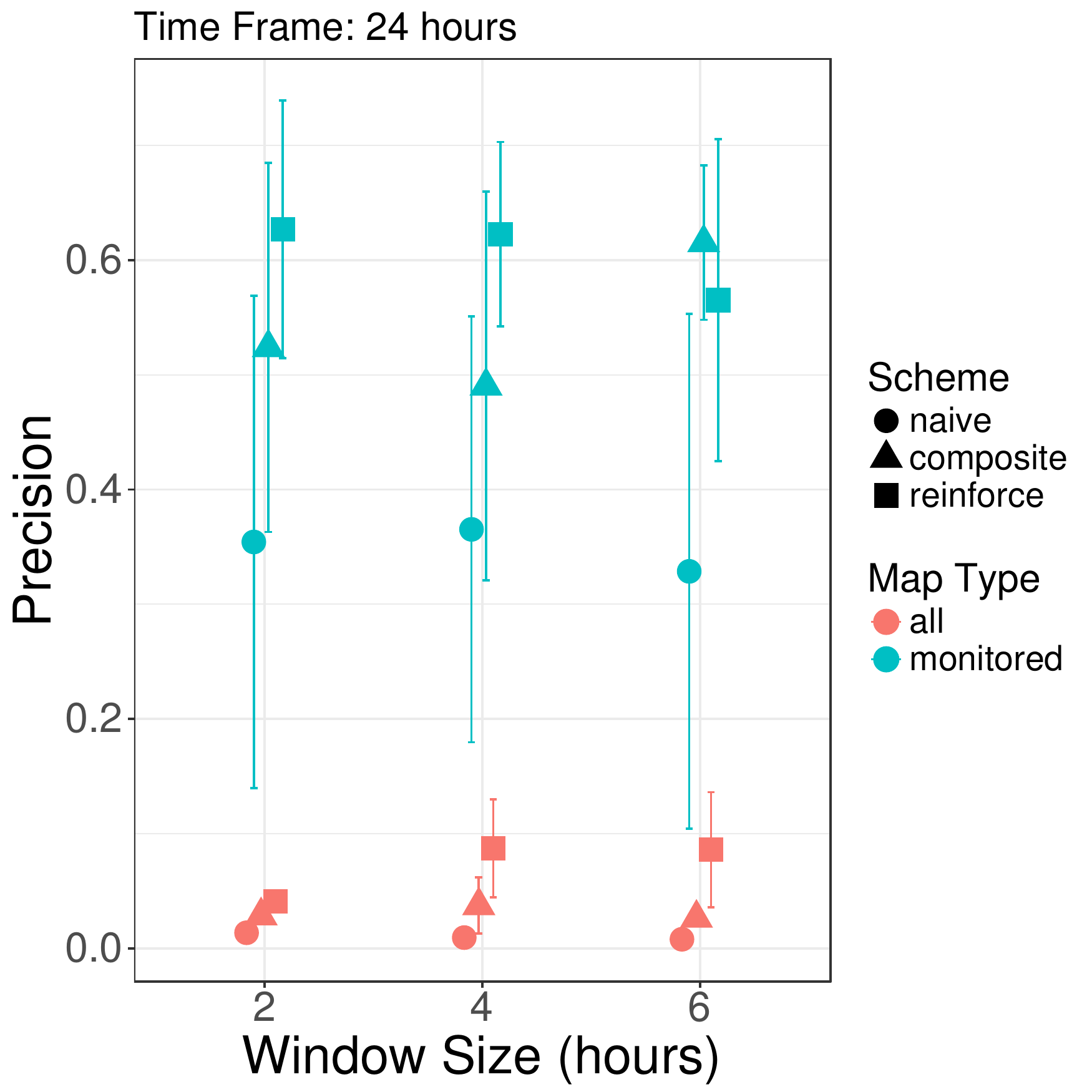}
	\caption{Precision}
	\label{fig:overallPrecisionMLA}
	\end{subfigure}
	\caption{Recall and Precision over the entire 24 hour period with the Louvain algorithm and the reinforcement schemes. The error bars represent 95\% confidence interval.}
\label{fig:overallRecallPrecisionMLA}
\end{figure}

The recall and precision for the Louvain algorithm are shown in~\fref{fig:overallRecallPrecisionMLA}. As with the recall within the time windows, the recall across time windows within a time frame is nearly complete for the scenario where only the sampled monitored botnet nodes are mapped to the nodes in the traffic traces that are active for the entire time frame. The precision for this case however is significantly higher for the reinforcement based schemes of composite and reinforced graphs. On the other hand, when all the botnet nodes are mapped to the always active nodes, as the worst case scenario, the recall for the composite and the reinforced graphs drops to about 95\% for the larger window sizes. The consistently high recall of the naive approach is explained by its extremely low precision, as among the tens of thousands of false positives the actual botnet nodes missed by the other schemes are also included in the detected P2P community. It is also worth noting that the confidence intervals reported across time windows are higher, partly because of the variations in precision across time windows, but also because of the smaller sample size because the experiments are repeated only 10 times for the entire time frame.

\begin{figure}[htbp]
\centering

	\begin{subfigure}[b]{0.49\columnwidth}
	\centering
	\includegraphics[width=\columnwidth]{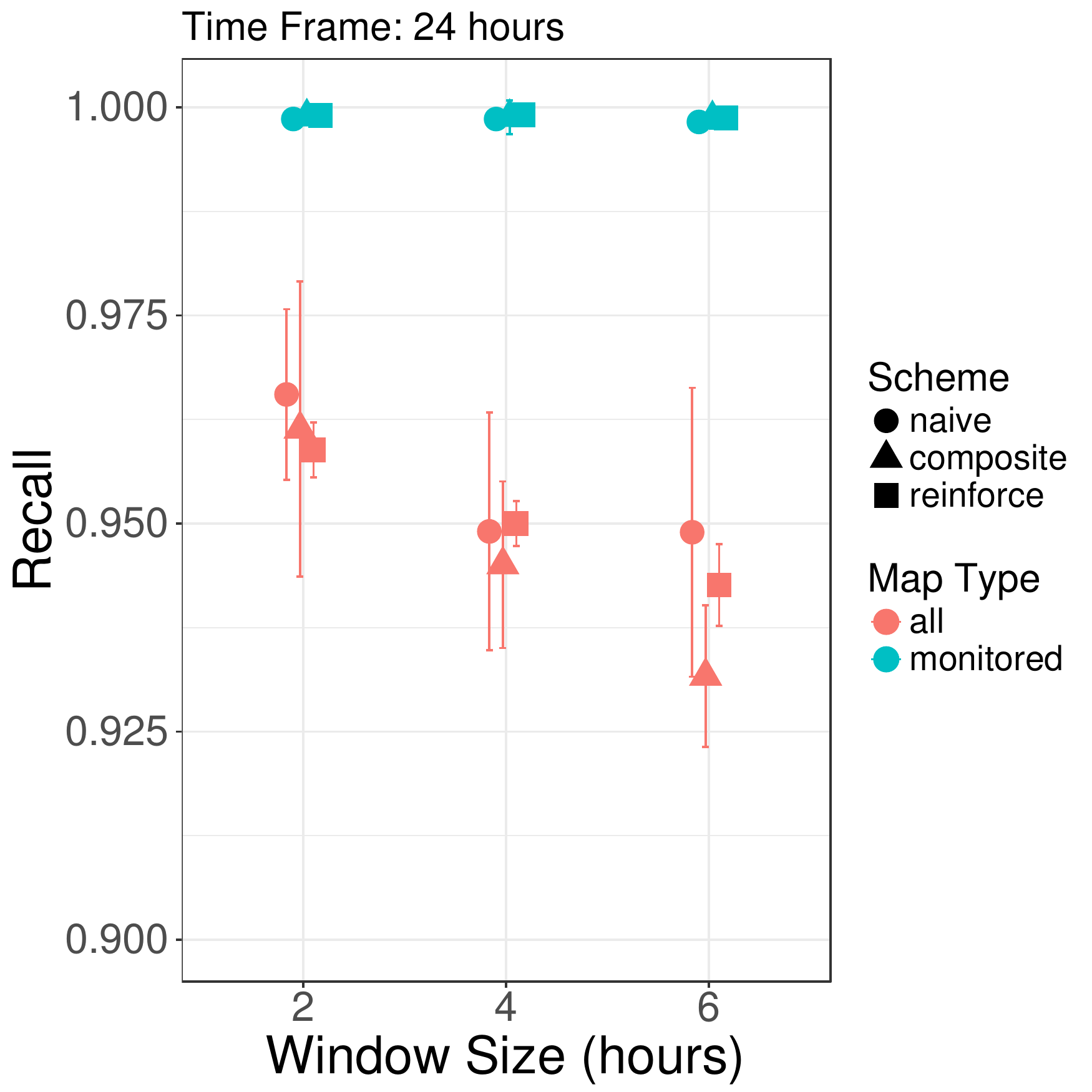}
	\caption{Recall}
	\label{fig:overallRecallLPA}
	\end{subfigure}%
	\hfill%
	\begin{subfigure}[b]{0.49\columnwidth}
	\centering
	\includegraphics[width=\columnwidth]{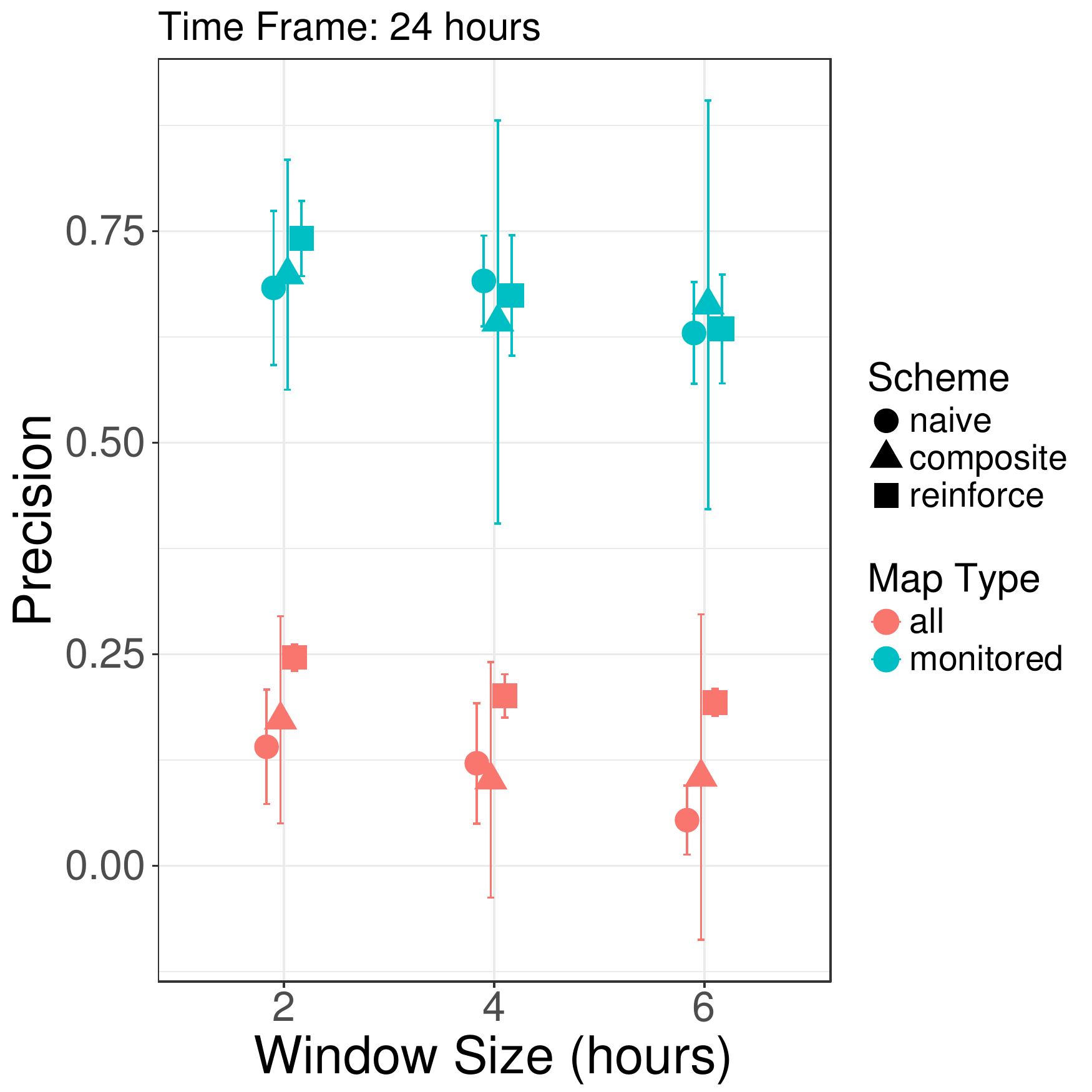}
	\caption{Precision}
	\label{fig:overallPrecisionLPA}
	\end{subfigure}
	\caption{Recall and Precision over the entire 24 hour period with the Label Propagation algorithm and the reinforcement schemes. The error bars represent 95\% confidence interval.}
\label{fig:overallRecallPrecisionLPA}
\end{figure}

The recall and precision for the Label Propagation algorithm (LPA) are shown in~\fref{fig:overallRecallPrecisionLPA}. Unlike the Louvain algorithm the LPA tends to detect smaller, more strongly connected communities. This results in higher precision within a time window. This advantage also carries over to the time frame when the detected P2P botnet communities are aggregated. The P2P community detected by the naive approach has between about 1 in 10 to 1 in 20 actual P2P botnet members, in the worst case. The reinforced graph approach is a significant improvement with about 1 in 4 to 1 in 5 actual P2P botnet members in the detected P2P community, with a slight (about 2\%) loss in recall for the worst case. In the best case, all three approaches have comparable recall and precision, though the reinforced graph approach has significantly less variability compared to the composite graph approach. 

\section{Conclusions}
In this paper we addressed the problem of detecting evolving P2P botnet communities in dynamic communication graphs. We proposed a discrete time model for building a sequence of communication graphs, and formulated the problem of P2P botnet detection in these dynamic graphs. Next we proposed a reinforcement approach to utilize the temporal information and the community structure information within each constituent time slice of a time window. We designed experiments for evaluating our proposed schemes with real-world dynamic botnet graphs combined with traffic traces from a large network. The results of the experiments show that the proposed reinforcement-based scheme improves precision not only for a particular time window, but also over a sequence of such time windows.

\lineskiplimit=0pt
\lineskip=0pt

\bibliographystyle{IEEEtran}
\bibliography{botcloud, zotero-lib}

\begin{thebibliography}{10}
\providecommand{\url}[1]{#1}
\csname url@samestyle\endcsname
\providecommand{\newblock}{\relax}
\providecommand{\bibinfo}[2]{#2}
\providecommand{\BIBentrySTDinterwordspacing}{\spaceskip=0pt\relax}
\providecommand{\BIBentryALTinterwordstretchfactor}{4}
\providecommand{\BIBentryALTinterwordspacing}{\spaceskip=\fontdimen2\font plus
\BIBentryALTinterwordstretchfactor\fontdimen3\font minus
  \fontdimen4\font\relax}
\providecommand{\BIBforeignlanguage}[2]{{%
\expandafter\ifx\csname l@#1\endcsname\relax
\typeout{** WARNING: IEEEtran.bst: No hyphenation pattern has been}%
\typeout{** loaded for the language `#1'. Using the pattern for}%
\typeout{** the default language instead.}%
\else
\language=\csname l@#1\endcsname
\fi
#2}}
\providecommand{\BIBdecl}{\relax}
\BIBdecl

\bibitem{zhuangPeerhunterDetectingPeertopeer2017}
D.~Zhuang and J.~M. Chang, ``Peerhunter: {{Detecting}} peer-to-peer botnets
  through community behavior analysis,'' in \emph{2017 {{IEEE Conference}} on
  {{Dependable}} and {{Secure Computing}}}.\hskip 1em plus 0.5em minus
  0.4em\relax {IEEE}, 2017, pp. 493--500.

\bibitem{joshiImprovedP2PBotnet2019}
H.~P. Joshi and R.~Dutta, ``Improved {{P2P Botnet Community Detection}}:
  {{Combining Modularity}} and {{Strong Community}},'' in \emph{2019 {{IEEE
  Global Communications Conference}} ({{GLOBECOM}})}, Dec. 2019, pp. 1--6.

\bibitem{nagarajaBotGrepFindingP2P2010}
S.~Nagaraja, P.~Mittal, C.-Y. Hong, M.~Caesar, and N.~Borisov, ``{{BotGrep}}:
  {{Finding P2P Bots}} with {{Structured Graph Analysis}}.'' \emph{USENIX
  Security Symposium}, vol.~10, 2010.

\bibitem{joshiGADFlyFastRobust2018}
H.~P. Joshi and R.~Dutta, ``{{GADFly}}: {{A Fast}} and {{Robust Algorithm}} to
  {{Detect P2P Botnets}} in {{Communication Graphs}},'' in \emph{2018 {{IEEE
  Global Communications Conference}} ({{GLOBECOM}})}, Dec. 2018, pp. 1--6.

\bibitem{coskunFriendsEnemyIdentifying2010}
B.~Coskun, S.~Dietrich, and N.~Memon, ``Friends of an enemy: Identifying local
  members of peer-to-peer botnets using mutual contacts,'' \emph{Proceedings of
  the 26th Annual Computer Security Applications Conference.}, 2010.

\bibitem{fortunatoCommunityDetectionGraphs2010}
S.~Fortunato, ``\BIBforeignlanguage{en}{Community detection in graphs},''
  \emph{\BIBforeignlanguage{en}{Physics Reports}}, vol. 486, no.~3, pp.
  75--174, Feb. 2010.

\bibitem{zanghiFastOnlineGraph2008}
H.~Zanghi, C.~Ambroise, and V.~Miele, ``\BIBforeignlanguage{en}{Fast online
  graph clustering via {{Erd\H{o}s}}\textendash{{R\'enyi}} mixture},''
  \emph{\BIBforeignlanguage{en}{Pattern Recognition}}, vol.~41, no.~12, pp.
  3592--3599, Dec. 2008.

\bibitem{peixotoInferringMesoscaleStructure2015}
T.~P. Peixoto, ``Inferring the mesoscale structure of layered, edge-valued, and
  time-varying networks,'' \emph{Physical Review E}, vol.~92, no.~4, p. 042807,
  Oct. 2015.

\bibitem{muchaCommunityStructureTimeDependent2010}
P.~J. Mucha, T.~Richardson, K.~Macon, M.~A. Porter, and J.-P. Onnela,
  ``\BIBforeignlanguage{en}{Community {{Structure}} in {{Time}}-{{Dependent}},
  {{Multiscale}}, and {{Multiplex Networks}}},''
  \emph{\BIBforeignlanguage{en}{Science}}, vol. 328, no. 5980, pp. 876--878,
  May 2010.

\bibitem{lancichinettiConsensusClusteringComplex2012}
A.~Lancichinetti and S.~Fortunato, ``\BIBforeignlanguage{en}{Consensus
  clustering in complex networks},'' \emph{\BIBforeignlanguage{en}{Scientific
  Reports}}, vol.~2, no.~1, p. 336, Mar. 2012.

\bibitem{haasResilienceP2PbasedBotnet2016}
S.~Haas, S.~Karuppayah, S.~Manickam, M.~M{\"u}hlh{\"a}user, and M.~Fischer,
  ``On the resilience of {{P2P}}-based botnet graphs,'' in \emph{2016 {{IEEE
  Conference}} on {{Communications}} and {{Network Security}} ({{CNS}})}, Oct.
  2016, pp. 225--233.

\bibitem{choTrafficDataRepository2000}
K.~Cho, K.~Mitsuya, and A.~Kato, ``Traffic data repository at the {{WIDE}}
  project,'' in \emph{Proceedings of the Annual Conference on {{USENIX Annual
  Technical Conference}}}, ser. {{ATEC}} '00.\hskip 1em plus 0.5em minus
  0.4em\relax {San Diego, California}: {USENIX Association}, Jun. 2000, p.~51.

\bibitem{analysisDayLifeInternet}
C.~C. f. A. I.~D. Analysis, ``A {{Day}} in the {{Life}} of the {{Internet}}
  ({{DITL}}),'' https://www.caida.org/projects/ditl/index.xml.

\bibitem{MAWIWorkingGroup}
``{{MAWI Working Group Traffic Archive}},'' http://mawi.wide.ad.jp/mawi/.

\bibitem{blondelFastUnfoldingCommunities2008}
V.~D. Blondel, J.-L. Guillaume, R.~Lambiotte, and E.~Lefebvre,
  ``\BIBforeignlanguage{en}{Fast unfolding of communities in large networks},''
  \emph{\BIBforeignlanguage{en}{Journal of Statistical Mechanics: Theory and
  Experiment}}, vol. 2008, no.~10, p. P10008, Oct. 2008.

\bibitem{raghavanLinearTimeAlgorithm2007}
U.~N. Raghavan, R.~Albert, and S.~Kumara, ``Near linear time algorithm to
  detect community structures in large-scale networks,'' \emph{Physical Review
  E}, vol.~76, p. 036106, 2007.

\end{thebibliography}

\end{document}